\documentstyle[12pt,epsf]{article}
\newcommand{\decentmargins}{
	\oddsidemargin=0in
	\evensidemargin=0in
	\textwidth=6.5in              

	\headheight=0pt
	\headsep=0pt
	\topmargin=0in
	\textheight=9.0in              

	\renewcommand{\baselinestretch}{1.5}
	}
\decentmargins
\begin{document}

\renewcommand{\small}{\normalsize} 
\newcommand{\sw}[1]{{\mbox{\scriptsize #1}}}
\catcode`@=11 
\def\lsim{\mathrel{\mathpalette\@versim<}}
\def\gsim{\mathrel{\mathpalette\@versim>}}
\def\gtlt{{\smash{\lower0.4ex\hbox{$<$}} \atop \smash{\raise0.4ex\hbox{$>$}}}}
\def\@versim#1#2{\lower0.2ex\vbox{\baselineskip\z@skip\lineskip\z@skip
  \lineskiplimit\z@\ialign{$\m@th#1\hfil##\hfil$\crcr#2\crcr\sim\crcr}}}
\catcode`@=12 
\newcommand{\shamelessUWplug}{%
\font\fortssbx=cmssbx10 scaled \magstep2
\hbox to \hsize{\vbox{\epsfysize=30pt\epsffile{uwlogo.eps}}
\quad\vbox to 30pt{\vfill\hbox{\fortssbx
University of Wisconsin - Madison}\vfill}\hfill}%
\vspace*{0.5in}
}

\begin{titlepage}
\shamelessUWplug
\begin{flushright}
	{MADPH-95-904}\\
        {CERN-TH/95-235}
\end{flushright}
\vspace{.25in}
\begin{center}
{\LARGE \bf Minijet activity in high $E_T$ jet events at the Tevatron}\\
\vspace{.5in}
{David Summers$^1$ and \ Dieter Zeppenfeld$^{1,2}$\\
\vspace{.25in}
$^1${\em Department of Physics, University of Wisconsin, Madison, WI 53706\\
$^2$CERN, 1211 Geneva 23, Switzerland}
}\\
\vspace{.25in}
ABSTRACT\\
\end{center}
\noindent Gluon bremsstrahlung in scattering events with high
transverse momentum jets is expected to increase markedly with the
hardness ($\sum E_T$) of the primary event. Within perturbative QCD we
estimate a probability of order unity to see additional minijets with
$E_T \gsim 15$~GeV in ``dijet'' events with $\sum E_T > 400$~GeV. The
veto of such minijets is a promising background rejection tool for the
Higgs search at the LHC.

\vspace{1in}
\begin{flushleft}
CERN-TH/95-235 \\
August 1995 \\
\end{flushleft}

\end{titlepage}

The study of detailed properties of high transverse energy ($E_T$) jet events
in $p\bar p$ collisions at the Tevatron has become an important testing ground
for perturbative QCD (pQCD)\cite{tdcs}. At the same time the investigation of
hadronization patterns in dijet events with widely separated jets has lead to
the discovery of rapidity gaps at the Tevatron~\cite{brandt} and has
demonstrated that the color structure of the hard scattering event has
dramatic consequences for the angular distribution of produced hadrons at
hadron colliders~\cite{colcoh,bjgap,fletcher}. More precisely it is the
$t$-channel exchange of color singlet quanta which can lead to rapidity gaps.
One of the most important members of this class of processes will be weak
boson scattering at hadron supercolliders, i.e. the electroweak process
$qq\to qqVV$, and, indeed, a rapidity gap trigger had been suggested as a
promising Higgs search tool at the SSC~\cite{bjgap}.

A modified rapidity gap trigger, in terms of minijets of $E_T>15-20$~GeV
instead of soft hadrons, was recently suggested for the study of weak boson
scattering at the LHC~\cite{bpz}. Due to the high luminosity of the LHC,
which leads to multiple $pp$ interactions in a single bunch crossing,
the search for rapidity regions without soft hadrons is rendered impractical.
Minijets of sufficiently high $E_T$, however, are unlikely to be produced
by multiple interactions.
On the theoretical side rapidity gaps are not well described by
pQCD which makes predictions about jets rather than individual
hadrons. Minijets, on the other hand, can be described by pQCD if
the perturbative expansion is taken far enough.

A key component of the minijet veto is the observation that the
emission of semihard partons, soft at the level of the hard scattering process
but still leading to distinct minijets, becomes quite probable in the very
hard QCD background processes to weak boson scattering. A tree level
calculation for the QCD $W$ pair background reveals that the production
cross sections for $WWj$ and $WWjj$ events become equal when the minimal
$E_T$ for which partons are identified as jets is taken to be
of order $30-40$~GeV~\cite{bpz}. This leads us to expect the
emission of multiple minijets with $E_T>20$~GeV in these background processes.
The $E_T$-scale can be understood qualitatively by noting that the invariant
mass of the two $W$'s must be in the 600--1000~GeV range to become a background
for a heavy Higgs boson and the presence of an additional forward tagging jet
increases the required c.m. energy $\sqrt{\hat s}$ into the TeV range. The
emission of additional gluons above a transverse momentum $E_T^\sw{min}$
is suppressed by a factor
$f_s \approx \alpha_s\;{\rm ln}\;(\hat s/{E_T^\sw{min}}^2)$
and this factor approaches unity for $E_T^\sw{min} = {\cal O}(30\;
{\rm GeV})$. Multiple minijet emission at
such large $E_T$'s should be observable even in a high luminosity environment
and it is the veto of these minijets which should substantially reduce the
backgrounds with little effect on the signal rates (due the different color
structure and QCD scales of the latter).

The approximate equality of $n$- and $n+1$-parton cross sections
clearly indicates that fixed order pQCD is no longer reliable
in this range. One would like to have experimental data on minijet
multiplicities, veto probabilities, angular distributions
and the like and to then gauge the pQCD calculations for new
physics backgrounds in the light of these direct observations. In this letter
we suggest that such information can already be obtained in the hardest QCD
events available now, namely in hard ``dijet'' events at the Tevatron.

In the following we consider pQCD predictions for dijet production
at the Tevatron in next-to-leading order (NLO), using the NLO Monte Carlo
program JETRAD of Giele, Glover and Kosower~\cite{gg}. MRSD$'_-$
distribution functions~\cite{MRSDmp} are used
and the strong coupling constant $\alpha_s(\mu)$ is evaluated at 2-loop order
with $\alpha_s(m_Z)=0.111$.  Both the renormalization and the factorization
scales are chosen as $\mu = \sum E_T/4$, where $\sum E_T$ is the scalar sum of
the parton transverse momenta. The program allows us to make accurate
estimates for the two-jet inclusive cross section ($\sigma_{2,\sw{incl}}$)
at the 1-loop level. In addition, three parton final states are simulated at
tree level which allows the study of soft minijet activity at leading order.
For the three parton cross section ($\sigma_3$) the usual scale ambiguity of
tree level calculations arises, which in the present case is exacerbated by the
very different scales of the hard event
($\sum E_T = {\cal O}\;(500 {\rm GeV})$) as
compared to the transverse momentum of the softest parton
($E_{T_{\scriptstyle 3}} = {\cal O}
(20$~GeV)). Truly we are dealing with a 2-scale problem and a rough estimate
of the ambiguities is obtained by comparing $\sigma_3(\mu=\sum E_T/4)$ with
$\sigma_3(\mu=\sum E_T/4) \cdot
\alpha_s(E_{T_{\scriptstyle 3}})/\alpha_s(\sum E_T/4)$
{\it i.e.} by effectively using the softest $E_T$
as the scale for emission of the additional parton. We will refer to these
two choices as $\mu_3 = \sum E_T/4$ and $\mu_3=E_{T_{\scriptstyle 3}}$,
respectively.

We cluster partons into jets using a D0 style cone algorithm~\cite{D0jet}.
If two partons are to be clustered into a jet we form the jet momentum as
\begin{eqnarray}
{\bf p}_\sw{jet} &=& {\bf p}_1 + {\bf p}_2 \; , \\
E_{T_\sw{jet}} &=& E_{T_{\scriptstyle 1}}+E_{T_{\scriptstyle 2}}
\; .
\end{eqnarray}
Each pair of partons is clustered into a temporary jet. Whenever both
partons are within a distance
\begin{equation}\label{delR}
\Delta R(\mbox{jet},\mbox{parton}) \equiv
\sqrt{(\eta_\sw{jet}-\eta_\sw{parton})^2
     +(\phi_\sw{jet}-\phi_\sw{parton})^2} < 0.7
\end{equation}
of the temporary jet, then these partons are clustered into an actual jet.
Since we are dealing with a maximum of three partons in the final state, at
most one such recombination of massless partons can occur in our program.

Final (clusters of) partons are then counted as jets if their transverse
energy $E_T$ and pseudorapidity $\eta$ satisfy the conditions
\begin{equation}\label{ETeta}
E_T > E_T^\sw{min}\, ,\qquad\qquad\qquad |\eta| < 3.5\; ,
\end{equation}
with, typically, $E_T^\sw{min}=15$~GeV. We define the $\sum E_T$ of
the event as the sum of the $E_T$'s of all observed jets.

\begin{figure}[t]
\epsfxsize=\hsize\epsffile{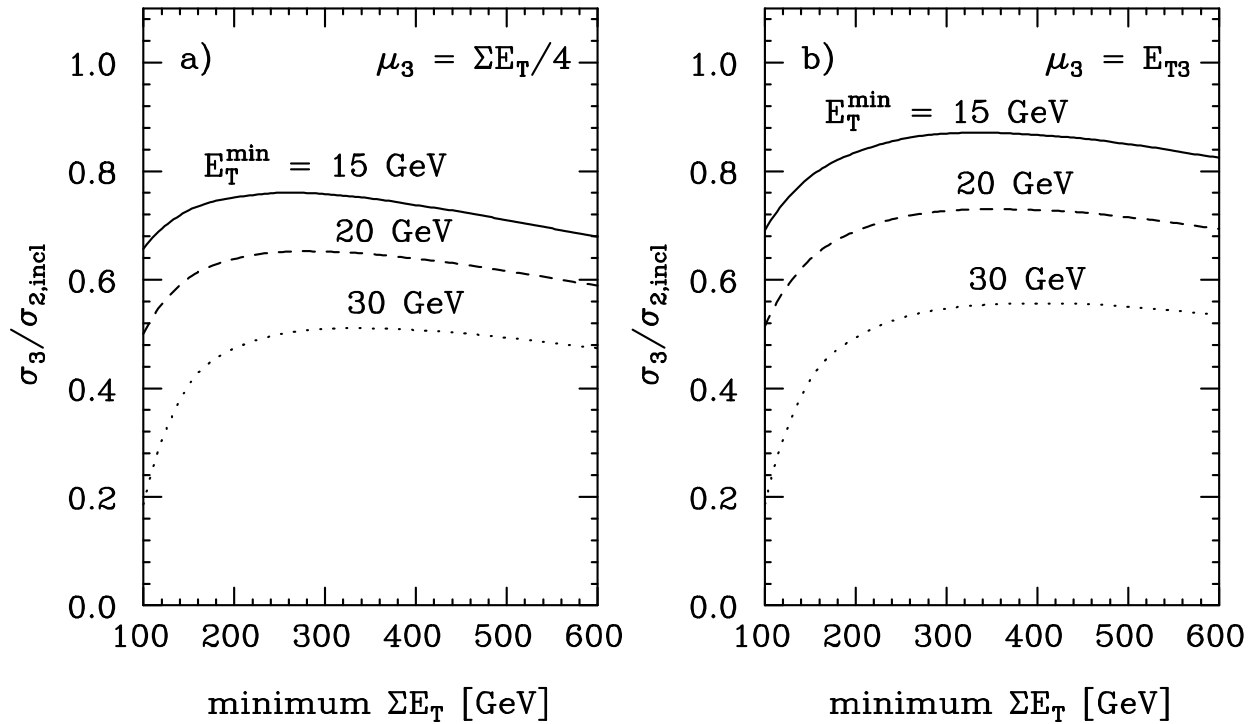}
\caption{
Fraction of 2-jet inclusive events that contain a third jet within the
acceptance requirements of Eqs.~\protect\ref{delR},\protect\ref{ETeta}.
The 3-jet cross
section is integrated above a minimal $\sum E_T$ and above minimal transverse
energies $E_{T_{\scriptstyle 3}}$ of the softest jet of 15~GeV (solid line),
20~GeV (dashed line) and 30~GeV (dotted line). The renormalization scale is
set to $\mu=\sum E_T/4$ in part a). Changing the effective coupling constant
for soft parton emission in the 3-jet cross section to
$\alpha_s(E_{T_{\scriptstyle 3}})$ yields the results shown in part b).
}
\protect\label{figsig23}
\end{figure}

In Fig.~\ref{figsig23} we show the fraction of 2 jet inclusive events that
contain a third jet, for values $E_T^\sw{min}= 15$, 20, and 30~GeV. The 3
jet fraction $f_3 = \sigma_3/\sigma_{2,\sw{incl}}$ is plotted as a function
of the $\sum E_T$ of the event and for the two choices of the effective
scale governing soft minijet emission. Below
$\approx 200$~GeV one observes a marked rise of the fraction of three jet
events with $\sum E_T$ which then flattens out around $\sum E_T=300$~GeV and
slowly decreases for larger values. This pattern can be understood by
considering the phase space, both in $E_T$ and pseudorapidity, which is
available to the third parton. By definition, this is the parton with the
smallest $E_T$ and hence its allowed range is $E_T^\sw{min}<E_T<\sum E_T/3$.
Integration of the partonic cross section $d\hat\sigma \sim dE_T/E_T$
over this range leads to a logarithmic increase of the 3 jet fraction and
explains the rise of $\sigma_3/\sigma_{2,\sw{incl}}$ for small $\sum E_T$.

\begin{figure}[t]
\hbox{\hfil\epsfxsize=0.6\hsize\epsffile{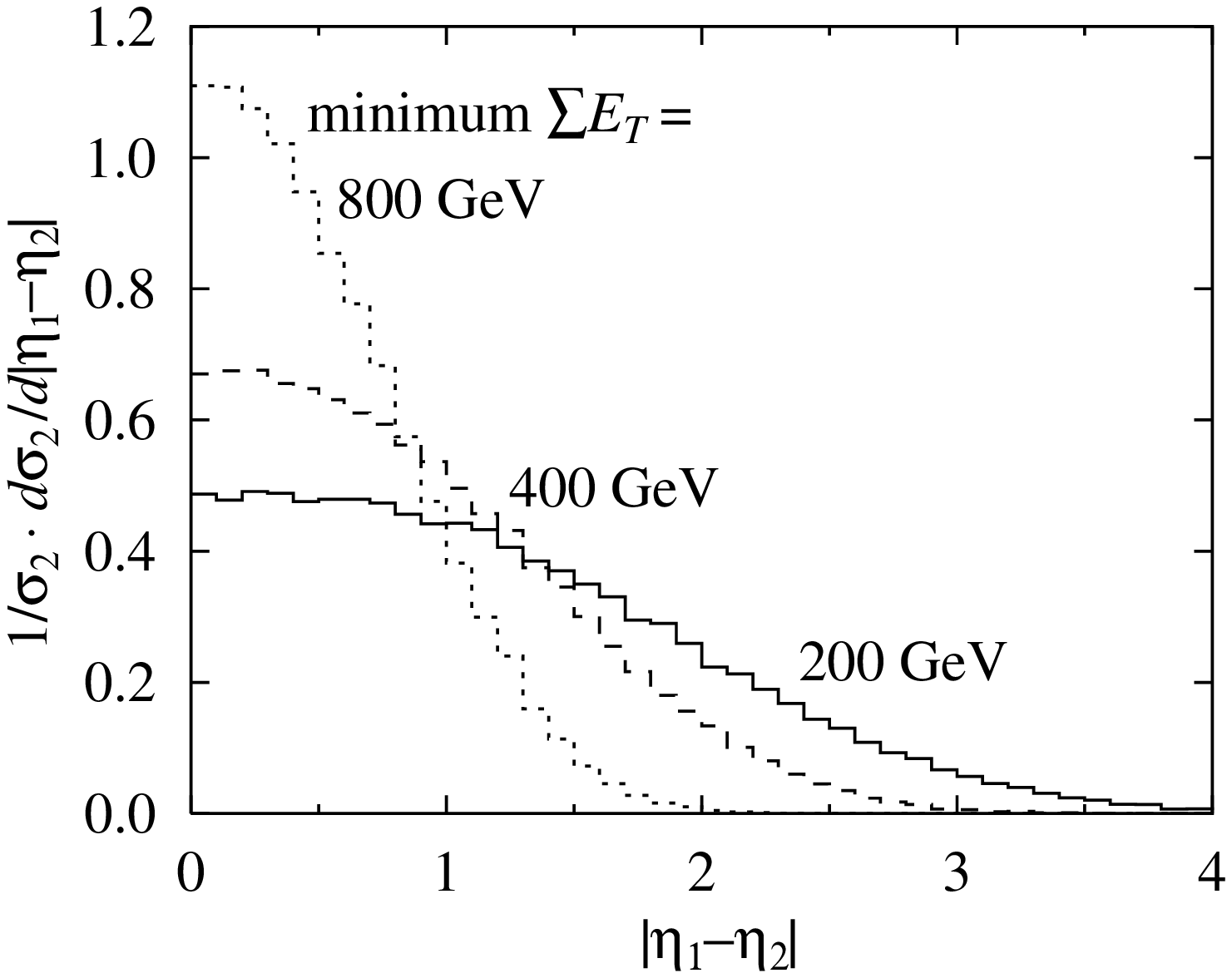}\hfil\hfil}
\caption{
Normalized rapidity distribution
$1/\sigma_2\;\cdot d\sigma_2/d|\eta_1-\eta_2|$
for 2-jet inclusive events at the Tevatron. Shown is the narrowing of the
rapidity range $|\eta_1-\eta_2|$ with increasing hardness ($\sum E_T$)
of the event.
}
\protect\label{figdeleta}
\end{figure}

For 2-jet events and, approximately, for 3-jet events with a soft central
jet the rapidity difference of the two highest $E_T$ jets and the c.m. energy
are related by
\begin{equation}
\sqrt{\hat s} = \sum E_T\cdot {\rm cosh}{\eta_1-\eta_2\over 2}\; .
\end{equation}
Due to the rapid decrease of parton luminosity with rising $\hat s$
only small rapidity differences $\eta_1-\eta_2$ can be reached at high
$\sum E_T$. This effect is demonstrated in Fig.~\ref{figdeleta}: the rapidity
distribution in the c.m. frame, $d\sigma/d|\eta_1-\eta_2|$ becomes narrower
with increasing $\sum E_T$. The emission of a third jet well outside the
rapidity interval set by the two hardest jets will markedly increase the
$\hat s$ of the event and hence will suffer from a lower parton luminosity.
Thus additional soft jets are predominantly produced in the rapidity range
set by the two hard jets and therefore the decrease of the average
$|\eta_1-\eta_2|$ with increasing $\sum E_T$ reduces the phase space for
emission of additional jets. This explains the decreasing fraction of three
jet events at large $\sum E_T$ which was observed in Fig.~\ref{figsig23}.

\begin{figure}[t]
\epsfxsize=\hsize\epsffile{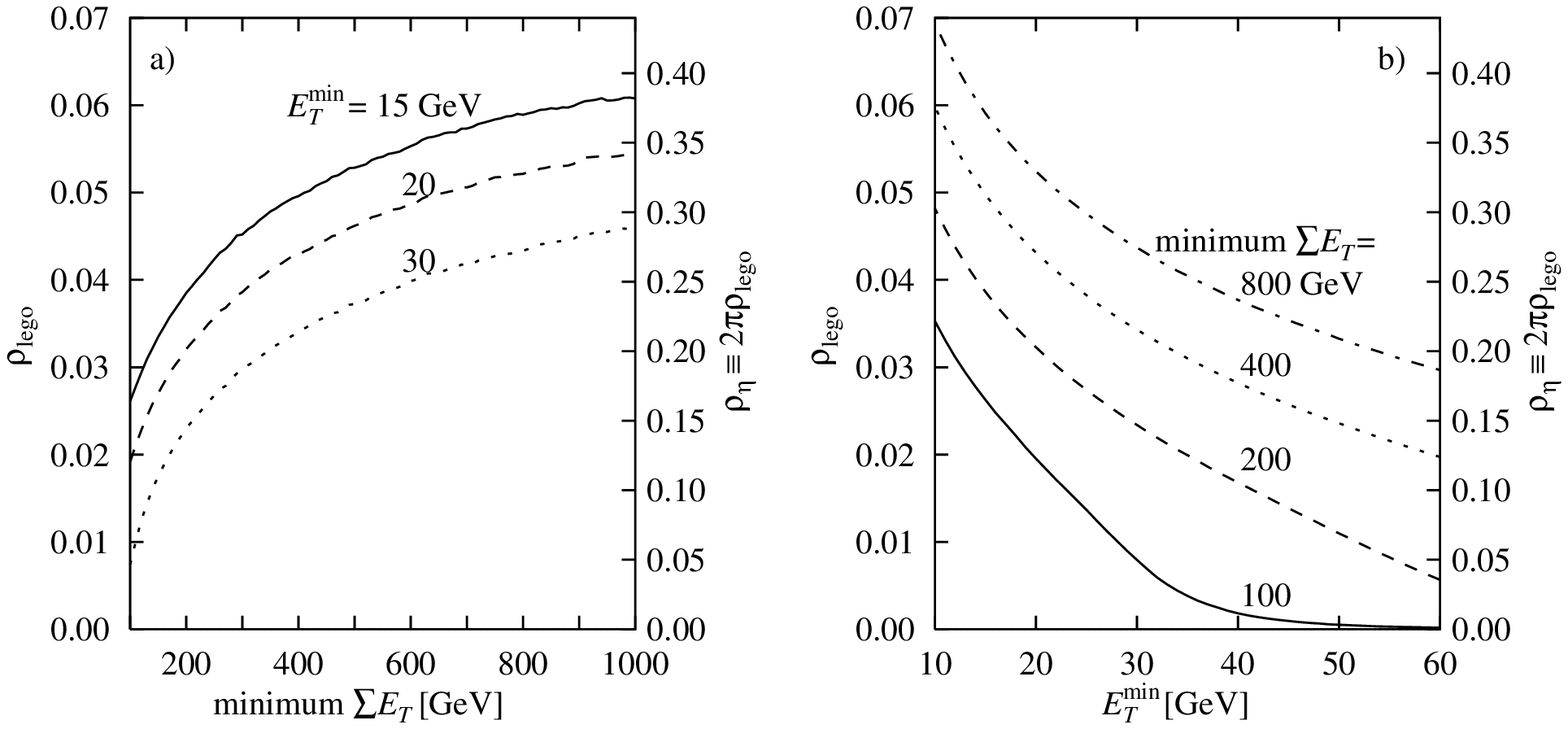}
\caption{
Lego-plot density $\rho_\protect\sw{lego}$ of soft jets between the two
hardest jets. Results are shown a) as a function of the minimal $\sum E_T$
and for three values of $E_T^\protect\sw{min}$ and b) as a function of
$E_T^\protect\sw{min}$  with $\sum E_T$ as a parameter.
The effective coupling
constant for soft parton emission in the 3-jet cross section has been set
to $\alpha_s(E_{T_{\scriptstyle 3}})$. On the r.h.s. the scale is multiplied
by a factor $2\pi$ to show the approximate density per unit of pseudorapidity.
}
\protect\label{figdens}
\end{figure}

These phase space effects are very process specific. In weak boson scattering
events at the LHC, for example, (and backgrounds to these events) the
available $\hat s$ is largely determined by the forward and backward tagging
jets and hence the kinematically favored rapidity range for minijet emission
is much larger than in dijet events at the Tevatron. For a comparison of
different processes one would like to eliminate these purely kinematical
effects. For jet production at the Tevatron our Monte Carlo studies show
that, to good precision, the probability for a 3rd jet to be produced in
the $\eta$ range between the two hardest jets rises linearly with the
$\eta$-$\phi$ phase space which is available for that third jet, that is
\begin{equation}
{\sigma_3( \eta_3 \in [\eta_1,\eta_2]) \over \sigma_{2,\sw{incl}}}
\sim A_\sw{lego} \equiv 2 \pi |\eta_1-\eta_2| - A_\sw{cone} \; ,
\end{equation}
where $A_\sw{cone}$ is the part of the lego-plot region between $\eta_1$ and
$\eta_2$ which is covered by the two cones of radius 0.7 about the two hardest
jets.
We therefore divide out the available phase space $A_\sw{lego}$ and
consider the probability density for emission of a third jet in the lego-plot.
In Fig.~\ref{figdens} we show the average density
\begin{equation}
\rho_\sw{lego} =
{1 \over \sigma_{2,\sw{incl}}} \; \int_{\eta_1}^{\eta_2} d\eta_3
\int_0^{2\pi} d\phi_3 \; {1\over A_\sw{lego}} \;
{d^2\sigma_3 \over d\eta_3 d\phi_3}
\end{equation}
for emission of the third, lowest $E_T$, jet in the interval
$[\eta_1, \eta_2]$ set by the two hardest jets.
The probability density as a function of the minimal $\sum E_T$ of the event
is shown in Fig.~\ref{figdens}a, while Fig.~\ref{figdens}b depicts the
same probability
density as a function of $E_T^\sw{min}$, the minimal transverse momentum
of the minijet. One now observes a monotonic increase with the hardness of the
overall event, $\sum E_T$. It is this increase which is characteristic of the
effect of enhanced minijet emission and which allows to distinguish it from
backgrounds due to the underlying event. Factorization of the hard event
implies that the background from the underlying event, that is minijet
emission caused by the interactions between the colored remnants of the
original $p \bar p$ pair, does not grow as the $\sum E_T$ of the hard event
increases. The minijet background from the underlying event may even decrease
at very high $\sum E_T$ because less longitudinal momentum is available for
the interactions between the $p \bar p$ remnants.

For very hard events, $\sum E_T > 400$~GeV say, the minijet emission
probability reaches remarkably large values, of order 0.3 per
unit of rapidity for minijets of $E_T > 15$~GeV.
These large probability densities reflect the approximate saturation
of the 2-jet inclusive cross section with 3-jet events which is evident in
Fig.~\ref{figsig23}. Clearly, for hard jet events at the Tevatron with
$\sum E_T \approx 400$~GeV the calculation of jet multiplicities within fixed
order pQCD is breaking down for jet thresholds as large as
$E_T^\sw{min}=30$~GeV. One should expect large probabilities for 4 or 5 jet
events as well, but a reliable prediction of these multijet fractions
goes beyond the perturbative tools at hand. A leading logarithm parton shower
calculation does allow to simulate such events. However, the scale of the
logarithms remains uncertain and the multijet rates are again subject to large
normalization uncertainties. Experimental input is needed in addition to
theoretical considerations. It should be noted that these events are not
rare, indeed $\sigma(\sum E_T>400\; {\rm GeV}) \approx 200$~pb.
Thus we expect a large
sample of high $\sum E_T$ events at the Tevatron which exhibit multiple
emission of additional minijets in the $10-30$~GeV $E_T$ range and which
can be used to study the characteristics of minijet emission.

Specific questions which should be addressed experimentally are
\begin{enumerate}
\item{}
What is the average multiplicity of minijets and how does it change with the
$\sum E_T$ of the event?

\item{}
How are minijet multiplicities distributed? The NLO Monte Carlo only allows to
generate events with 0 and 1 minijet.

\item{}
Down to which $E_T^\sw{min}$ can one distinguish minijet activity arising
in the hard scattering event from minijets in the underlying event?

\item{}
What is the probability for zero minijet emission in a hard scattering
event? How does this probability change with $\sum E_T$? How
does it depend on the minimal transverse energy $E_T^\sw{min}$ which is
required for a cluster of hadrons to be identified as a jet?


\item
Does the effective coupling, which governs the emission of minijets, depend
on the hard scattering scale, the soft scale, or some combination of the two?

\end{enumerate}

The results shown in Fig.~\ref{figsig23} can only be considered crude
estimates, as can be seen from the differing predictions when changing the
scale of soft emission from $\sum E_T/4$ to $E_{T_{\scriptstyle 3}}$.
Theoretically this question of scale can be resolved by going to the next
order in pQCD, in this case calculating 3 jet production at NLO.
Experimental information on this scale is obtained by comparing the rise
of the lego-plot density as a function of $\sum E_T$ with the
$\mu_3=E_{T_{\scriptstyle 3}}$ prediction in Fig.~\ref{figdens}a.
Considering the large uncertainties of our pQCD calculation it is well
possible, however, that a relatively large correction factor $F$ is needed
to relate the experimental minijet density with the pQCD expectation,
\begin{equation}
\mbox{Minijet density} = F\,\rho_\sw{lego} =
{F \over \sigma_{2,\sw{incl}}}  \; \int_{\eta_1}^{\eta_2} d\eta_3
\int_0^{2\pi} d\phi_3 \; {1\over A_\sw{lego}} \;
{d^2\sigma_3 \over d\eta_3 d\phi_3} \; .
\end{equation}
Measuring the dependence of this correction factor on the kinematics of the
hard event ($\sum E_T$ and $|\eta_1-\eta_2|$) would provide a useful guide
for theoretical
improvements, which would then help to more reliably predict minijet emission
probabilities in hard scattering events at the LHC.

\section*{Acknowledgements}
We thank E.~W.~N.~Glover for making the NLO dijet Monte Carlo program JETRAD
available to us, and for helpful discussions. Stimulating discussions on jet
physics with J. Blazey, E. Gallas and Ki Suk Hahn are greatfully acknowledged.
DJS would like to thank Rutherford Appleton Particle Physics Theory group for
its kind hospitality, where the finishing touches were made to this work.
This research was supported in part by the University of Wisconsin Research
Committee with funds granted by the Wisconsin Alumni Research Foundation and
in part by the U.~S.~Department of Energy under Contract No.~DE-AC02-76ER00881.

\newpage             


\end{document}